\documentclass[12pt]{article}

\usepackage{fullpage}
\usepackage{cite}
\usepackage{graphicx}
\usepackage{amsmath}
\usepackage{amssymb}

\allowdisplaybreaks
\title{}
\date{}

\def\beq{\begin{equation}}
\def\eeq{\end{equation}}
\def\beqa{\begin{eqnarray}}
\def\eeqa{\end{eqnarray}}
\def\eq#1{Eq.~(\ref{#1})}

\newcommand{\secn}[1]{Section~\ref{#1}}

\newcommand{\be}{\begin{equation}}
\newcommand{\ee}{\end{equation}}
\newcommand{\bea}{\begin{eqnarray}}
\newcommand{\eea}{\end{eqnarray}}
\newcommand{\nn}{\nonumber}

\newcommand{\Tr}{\mbox{Tr}}
\newcommand{\ord}{{\cal O}}
\newcommand{\as}{\alpha_s}
\newcommand{\eps}{\epsilon}

\def\slash#1{#1 \hskip-0.45em /}

\begin{document}
\bibliographystyle{utphys}


\titlepage

\begin{flushright}
Edinburgh 2014/18 \\
NIKHEF/2014-039\\
ITF-UU-14/24
\end{flushright}

\vspace*{1.2cm}

\begin{center}
{\Large \bf The method of regions and next-to-soft corrections in Drell-Yan production}

\vspace*{1cm} \textsc{D. Bonocore$^a$,
  E. Laenen$^{a,b,c}$,
  L. Magnea$^{d}$,
  L. Vernazza$^{d,e}$ and
  C. D. White$^f$} \\

\vspace*{1.5cm} 

$^a$ Nikhef, Science Park 105, NL-1098 XG Amsterdam, The Netherlands

\vspace*{0.3cm} 

$^b$ ITFA, University of Amsterdam, Science Park 904, Amsterdam, The Netherlands

\vspace*{0.3cm} 

$^c$ ITF, Utrecht University, Leuvenlaan 4, Utrecht, The Netherlands

\vspace*{0.3cm} 

$^d$ Dipartimento di Fisica Teorica, Universit\`a di Torino, and \\
INFN, Sezione di Torino, Via P. Giuria 1, I-10125 Torino, Italy

\vspace*{0.3cm} 

$^e$ Higgs Centre for Theoretical Physics, School of Physics and Astronomy, \\ 
The University of Edinburgh, Edinburgh EH9 3JZ, Scotland, UK

\vspace*{0.3cm} 

$^f$ SUPA, School of Physics and Astronomy, University of Glasgow,\\ 
Glasgow G12 8QQ, Scotland, UK

\end{center}

\vspace*{1cm}

\begin{abstract}
\noindent
We perform a case study of the behavior of gluon radiation beyond the
soft approximation, using as an example the Drell-Yan production cross
section at NNLO. We draw a careful distinction between the eikonal
expansion, which is in powers of the soft gluon energies, and the
expansion in powers of the threshold variable $1 - z$, which involves
important hard-collinear effects. Focusing on the contribution to the
NNLO Drell-Yan K-factor arising from real-virtual interference, we use
the method of regions to classify all relevant contributions up to
next-to-leading power in the threshold expansion.  With this method,
we reproduce the exact two-loop result to the required accuracy,
including $z$-independent non-logarithmic contributions, and we
precisely identify the origin of the soft-collinear interference which
breaks simple soft-gluon factorization at next-to-eikonal level. Our
results pave the way for the development of a general factorisation
formula for next-to-leading-power threshold logarithms, and clarify
the nature of loop corrections to a set of recently proposed
next-to-soft theorems.
\end{abstract}


\section{Introduction}
\label{intro}

It is well known that singularities arise in perturbative scattering amplitudes 
due to low-energy (soft) emission of massless gauge bosons, and to collinear 
splittings of massless particles. These {\it infrared} (IR) singularities cancel 
for suitably defined inclusive cross sections, once real and virtual diagrams 
are combined~\cite{Bloch:1937pw}; more generally, they are known to factorize 
at the level of scattering amplitudes~\cite{Dixon:2008gr}, and their general 
structure in the case of multi-parton non-abelian gauge amplitudes has 
been the subject of much recent activity (for a recent summary, see for 
example~\cite{Gardi:2014kpa,Magnea:2014vha}, and references therein).

Even for finite, infrared-safe cross sections, residual contributions persist after 
the cancellation of singularities, taking the form of potentially large kinematic 
logarithms at all orders in perturbation theory, which in general need to be 
resummed. In the generic case of multi-scale processes, these logarithms
can have a variety of arguments, such as transverse momenta which vanish
at Born level, or event shape variables which vanish in the two-jet limit. In this 
note, we will concentrate on {\it threshold} logarithms, which arise in inclusive 
cross sections when real radiation is forced to be soft or collinear by the properties
of the selected final state. Examples are: electroweak annihilation processes,
such as Drell-Yan production or Higgs production via gluon fusion, where the
threshold variables are $z = Q^2/\hat{s}$ and $z = M_H^2/\hat{s}$, respectively,
with $\hat{s}$ the partonic center-of-mass energy; Deep Inelastic Scattering (DIS),
where the threshold variable is the partonic version of Bjorken $x$; and $t \bar{t}$
production, where the threshold variable is $z = 4 m_t^2/\hat{s}$. In all of these 
cases the cancellation of infrared singularities leaves behind logarithms of the
general form $\alpha_s^n \, (1 - z)^m \log^p (1 - z)$, with $0 \leq p \leq 2 n - 1$, 
and $m \geq -1$.

Contributions with $m = - 1$, which we describe as {\it leading power} (LP) 
threshold logarithms, have been extensively studied, and successfully 
resummed to very high logarithmic accuracy using a variety of 
formalisms~\cite{Sterman:1986aj,Catani:1989ne,Korchemsky:1992xv,
Contopanagos:1996nh,Forte:2002ni,Becher:2006nr}. It is however known 
that also logarithms accompanied by subleading powers of the threshold 
variable, most notably those with $m = 0$, which we call {\it next-to-leading 
power} (NLP) threshold logarithms, can give numerically significant 
contributions~\cite{Kramer:1996iq}. In recent years, a number of studies 
have appeared~\cite{Dokshitzer:2005bf,Laenen:2008gt,Laenen:2008ux,
Grunberg:2009yi,Laenen:2010uz,Almasy:2010wn,Ball:2013bra,Altinoluk:2014oxa,
Apolinario:2014csa,deFlorian:2014vta} developing our understanding of
certain classes of NLP threshold logarithms. A full-fledged resummation 
formalism for NLP logarithms is however still not available.

An important class of NLP threshold logarithms, which is the best
studied so far, is generated by contributions to scattering amplitudes
that arise from the emission of soft gluons, at next-to-leading power
in the soft gluon energy. We call these contributions {\it
  next-to-eikonal} (NE), or {\it next-to-soft}. It has been known for
many years, at least in the abelian
case~\cite{Low:1958sn,Burnett:1967km, DelDuca:1990gz}, that
next-to-soft emissions share many of the universal features that
characterize leading-power soft radiation, which is described by the
eikonal approximation. This understanding, to some extent, has been
generalized to non-abelian theories
in~\cite{Laenen:2008gt,Gardi:2010rn,Laenen:2010uz}, where it was shown
that the eikonal approximation can be generalized to take into account
next-to-soft effects, while preserving many of the nice universality
and factorization properties which are present at leading
power. Ultimately, however, in order to organize all NLP threshold
logarithms, one must include also the effects of collinear
emissions. The importance of collinear emissions is evident in the
case of processes with final-state jets, for example DIS, where some
threshold logarithms are directly associated with the mass of the
current jet. It is crucial to realize, however, that collinear
emissions will also contribute to NLP logarithms for processes, like
Drell-Yan or Higgs production, where real radiation is forced to be
soft by phase space constraints. In such cases the soft expansion
breaks down because singularities arising from virtual hard collinear
gluons interfere with the soft approximation.  This issue was first
tackled, in the abelian case, in Ref.~\cite{DelDuca:1990gz}, and
similar effects were noted in
Refs.~\cite{Laenen:2008gt,Laenen:2008ux}. The analysis of the present
paper will precisely identify the origin of these interfering
contributions in an example involving real-virtual interference for
the Drell-Yan cross section at NNLO.

Quite interestingly, next-to-soft corrections to scattering amplitudes
have been the focus of intense recent research also from a more formal
point of view. It is well known that leading-power soft radiation can
be studied with eikonal methods both in gauge theories and in
gravity~\cite{Weinberg:1965nx,Naculich:2011ry,White:2011yy,Akhoury:2011kq,
  Beneke:2012xa}. Recently, Ref.~\cite{Cachazo:2014fwa} conjectured
that next-to-soft behaviour at tree-level is universal in gravity,
based on the observation that the known universal soft
behaviour~\cite{Weinberg:1965nx} can be obtained via a Ward identity
associated with the Bondi-Metzner-Sachs (BMS) symmetry at past and
future null
infinity~\cite{He:2014laa}. Reference~\cite{Casali:2014xpa}
generalised this to Yang-Mills theory, and there have been a number of
follow-up studies~\cite{Schwab:2014xua,
  Larkoski:2014hta,Kapec:2014opa,Geyer:2014lca,Schwab:2014fia,Bianchi:2014gla,
  Afkhami-Jeddi:2014fia,Adamo:2014yya,Bern:2014oka,Bern:2014vva,Broedel:2014fsa,He:2014bga,Cachazo:2014dia}.
In particular, Ref.~\cite{White:2014qia} pointed out the relationship
between this body of work and the more phenomenological results of
Refs.~\cite{Low:1958sn,Burnett:1967km,
  DelDuca:1990gz,Laenen:2008gt,Laenen:2010uz}. A key point of
contention in the current literature is whether next-to-soft theorems
receive corrections at loop level. As Ref.~\cite{Cachazo:2014dia}
makes clear, this is related to the sequential order in which the
expansions in soft momentum and the dimensional regularisation
parameter $\epsilon$ (in $4 - 2 \epsilon$ dimensions) are carried
out. The authors of Ref.~\cite{Cachazo:2014dia} state that the soft
expansion should be carried out first (with $\epsilon$ kept
non-zero). Loop corrections were further explored in
Refs.~\cite{Bianchi:2014gla,Bern:2014oka,He:2014bga}, with
Ref.~\cite{Bern:2014oka} advocating that the soft expansion be carried
out {\it after} the $\epsilon$-expansion, which would correspond to
how complete amplitudes are usually calculated.

Our aim in this letter is to perform a case study of NLP threshold logarithms at loop 
level in Drell-Yan production, including in particular those that originate from next-to-soft 
corrections to the corresponding scattering amplitude. There are a number of motivations 
for doing so. First, our ultimate aim (building on the work of refs.~\cite{Laenen:2008gt,
Laenen:2010uz}), is to develop a fully general resummation prescription for NLP threshold
logarithms. Our investigation here will provide crucial data in this regard, although we
postpone a detailed discussion of factorisation at NLP accuracy  to a subsequent 
paper~\cite{NEinprep}. Secondly, by explicitly characterising contributions in Drell-Yan 
according to their soft and/or collinear behaviour, we will be able to concretely examine 
the issue of loop corrections to next-to-soft theorems, including the interplay between 
the dimensional regularisation and soft expansions. We will verify explicitly that performing 
the $\epsilon$ expansion {\it before} the soft expansion correctly reproduces known results 
that are sensitive to this ordering. The reason is, as might be expected, the fact that there
are collinear singularities arising from virtual exchanges of hard collinear gluons, which
are not correctly taken into account if one performs a soft expansion before loop
integrations.

More specifically, we will examine the $K$-factor for Drell-Yan
production at NNLO, concentrating on those terms which arise from
having one real and one virtual gluon emission, which are ideally
suited to examine the questions posed above. Indeed, logarithms
arising from double real emission were already understood from an
effective next-to-soft approach in Ref.~\cite{Laenen:2010uz}, using
the fact that, for electroweak annihilation processes, real radiation
near threshold is forced to be soft. Double virtual corrections, on
the other hand, have a trivial dependence on the threshold variable
$z$, and do not influence the present considerations. In this letter,
we will further concentrate on terms proportional to the colour
prefactor $C_F^2$, which are the same as those that would be obtained
in an abelian theory, as considered in the earlier work
of~\cite{Low:1958sn,Burnett:1967km,DelDuca:1990gz}. This is sufficient
to illustrate our main points, and a complete analysis will be given
in forthcoming work~\cite{NEinprep}.  Our task here will be to perform
a detailed momentum-space analysis of the selected contributions, and
trace the origin of all NLP threshold logarithms to hard, soft, or
collinear configurations. To this end, we will use the {\it method of
  regions}, as developed in~\cite{Beneke:1997zp}. A similar analysis
has recently been performed in the case of Higgs production in gluon
fusion, to an impressive N$^3$LO accuracy~\cite{Anastasiou:2013mca},
as part of the complete calculation of the soft and virtual
contributions to the cross section at this order.  In that case, the
method of regions was used as an alternative technique to check the
validity of the threshold expansion, and as a method to investigate
the convergence properties of the expansion itself~\footnote{For a
  discussion of the limits of the threshold expansion in this process,
  see ref.~\cite{Herzog:2014wja}.}. Our goal is different, namely to
analyse the factorisation properties of various diagrammatic
contributions to the cross section.  As a consequence, in
Ref.~\cite{Anastasiou:2013mca} the method of regions was applied after
reduction to scalar master integrals, while here we apply it to
complete diagrams, thus making it easier to trace various sources of
next-to-soft behaviour in our chosen (Feynman) gauge. Furthermore, for
the specific NNLO contributions we focus on, we will be able to show
how the method of regions gives an exact account of threshold
contributions also at next-to-leading power.

Our results will prove useful in the development of a factorisation
formula for NLP threshold logarithms, which will generalise the
well-known soft-collinear factorisation formula at leading power (see,
for example, Ref.~\cite{Gardi:2009zv} for a review of the latter);
work in this direction is in
progress~\cite{NEinprep}\footnote{Progress can also be made using
  effective field theory techniques~\cite{Duffinprep}. One of the
  authors (CDW) is very grateful to Duff Neill for correspondence on
  this point, including sharing an early draft of
  Ref.~\cite{Duffinprep}.}. Interestingly, we find that our analysis
with the method of regions is able to reproduce correctly all NLP
threshold corrections, including terms with $m = 0$ and $p = 0$, which
have no logarithms at all, and would correspond to terms of order
$1/N$ in a Mellin-space analysis, with no $\log N$ enhancements. We
think this gives evidence for the existence of a systematic
organization of threshold contributions to cross sections, order by
order in $m$.

The structure of the letter is as follows. In \secn{sec:DY} we review necessary 
information about Drell-Yan production. In \secn{sec:calc} we apply the method of 
regions to classify all abelian-like terms in the real-virtual contribution to the NNLO
Drell-Yan $K$-factor. In \secn{sec:loop}, we interpret our results in light of
Refs.~\cite{Cachazo:2014dia,Bianchi:2014gla,Bern:2014oka,He:2014bga},
focusing in particular on the required ordering of the soft and $\epsilon$-expansions. 
We discuss our results and conclude in \secn{sec:conclude}.


\section{Real-Virtual interference in Drell-Yan at NNLO}
\label{sec:DY}

As discussed in the introduction, we consider Drell-Yan production of a virtual 
vector boson~\cite{Drell:1970wh}, which at leading order proceeds via the process
\beq
  q(p) + \bar{q} (\bar{p}) \, \rightarrow \, V^*(Q) \, ,
\label{DYLO}
\eeq
where we do not display flavor indices, so that the vector boson $V$ could be a photon, a 
$Z$ or a $W^\pm$ boson. The threshold variable in this case is $z = Q^2/\hat{s}$,
with $Q = p + \bar{p}$. The Drell-Yan $K$-factor at ${\cal O}(\alpha_s^n)$ is defined by 
\beq
  K^{(n)}(z) \, = \, \frac{1}{\sigma^{(0)}} \, \frac{d \sigma^{(n)}(z)}{dz} \, ,
\label{Kndef}
\eeq
where $\sigma^{(n)}$ is the total cross-section including terms up to ${\cal O}(\alpha_s^n)$. 
The cross section has been calculated exactly up to $n = 2$ in Refs~\cite{Altarelli:1979ub,
Hamberg:1990np,Matsuura:1988sm,Matsuura:1988nd}, which allows scrutiny of threshold 
logarithms both at LP and at NLP accuracy. The relevant contributions take the form
\beq
  {\rm LP:}  \quad  \alpha_s^n \left[ \frac{\log^m(1 - z)}{1 - z} \right]_+ \equiv 
  \alpha_s^n \, {\cal D}_m(z) \, ; \quad {\rm NLP:} \quad 
  \alpha_s^n \log^m (1- z) \, , \quad 0 \leq m \leq 2 n - 1 \, .
\label{logs}
\eeq
Leading power logarithms, supplemented by terms proportional to $\delta(1 - z)$, form the
so-called `soft $+$ virtual' contribution, which has been recently computed to N$^3$LO in
Ref.~\cite{Anastasiou:2013mca}. We see that NLP contributions show up as pure logarithms,
integrably singular in the threshold region $z \rightarrow 1$. At NLO, such terms arise only 
through real emission contributions: these were analysed in Ref.~\cite{Laenen:2010uz}, 
together with the double real emission contributions at NNLO, and shown to be reproducible 
from an effective next-to-eikonal approach. This is due to a lack of contamination in 
tree-level DY production from hard collinear effects, which is not true in more generic 
processes, or at loop level: beyond NLO, also for Drell-Yan kinematics, one must then 
differentiate between the expansion in emitted (soft) gluon momentum, and the {\it threshold} 
expansion which also includes collinear effects. 

\begin{figure}
\begin{center}
\scalebox{0.70}{\includegraphics{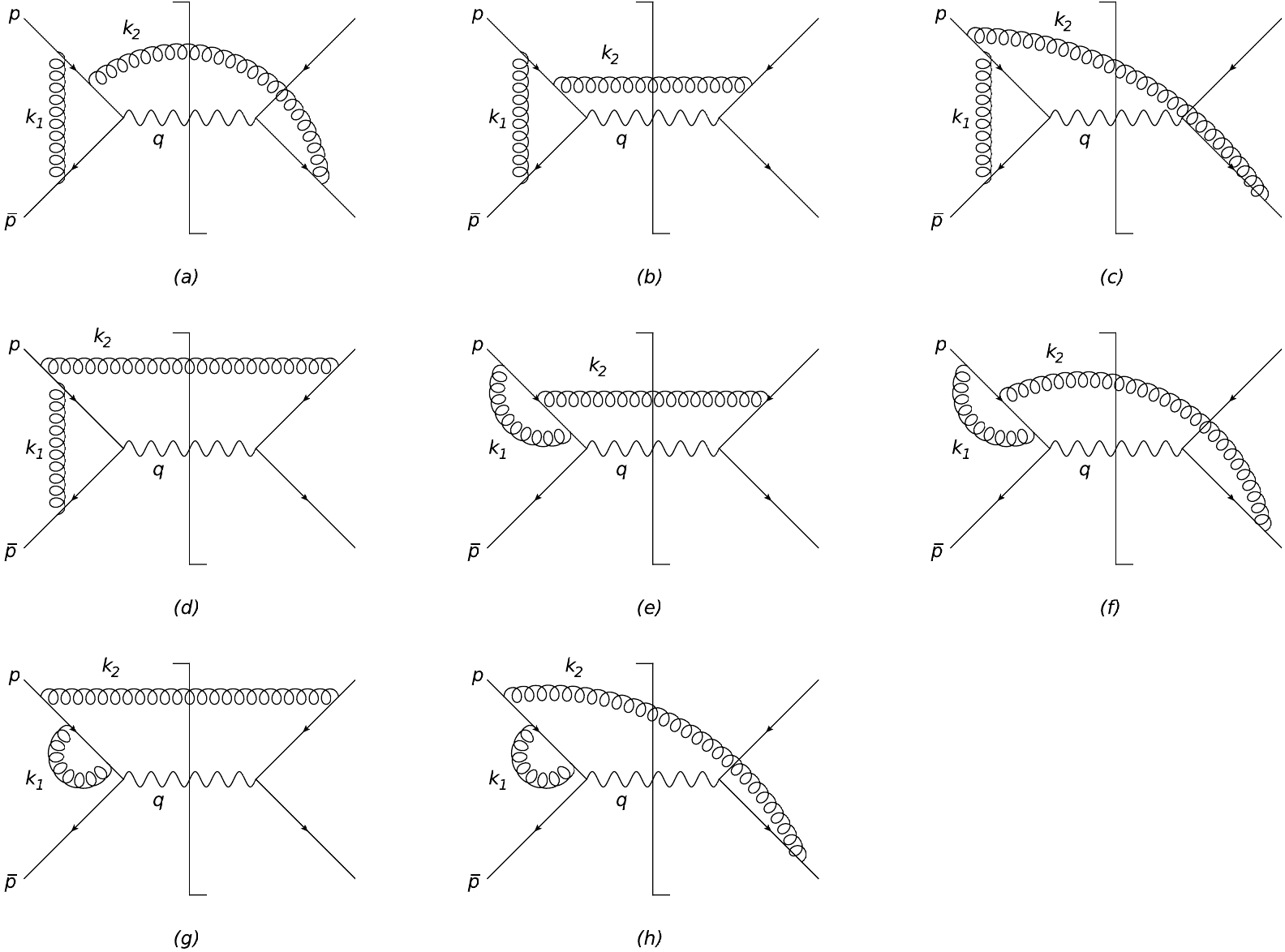}}
\caption{Abelian-like cut diagrams contributing to the Drell-Yan cross section at NNLO,
  involving one real and one virtual gluon. Diagrams obtained by interchanging 
  $p \leftrightarrow \bar{p}$ and/or complex conjugation are not shown.}
\label{sigbdiags}
\end{center}
\end{figure}

Following on from Ref.~\cite{Laenen:2008gt}, the next milestone in
understanding the structure of NLP threshold logs is to examine
one-loop graphs at NNLO, involving one real and one virtual
gluon. These were not considered explicitly in
Refs.~\cite{Laenen:2008gt, Laenen:2010uz}, due to the fact that hard
collinear singularities were not accounted for.  The interplay between
(next-to) soft and collinear effects has been discussed at length in
Ref.~\cite{DelDuca:1990gz}, at the price of neglecting discussion of
double counting issues between gluon emissions that are simultaneously
soft and collinear. In order to clarify these issues, we concentrate
here on the abelian-like contribution to the NNLO real-virtual
interference contributions to the $K$-factor, corresponding to the
(cut) Feynman diagrams shown in Fig~\ref{sigbdiags}.
As an example, to fix our notation, we note that diagram (a) contributes
\beqa
\label{NNLO3}
  F_a (z)  & = & g_s^4 \int[d k_1] \, [d k_2] \,
  ( 2 \pi ) \, \delta(k^2_2) \, \theta(k^0_2) \, \delta \left( \frac{\omega}{2} - k^0_2 \right) \, 
  \frac{1}{k_1^2} \\
  & & \times \, \Tr \bigg[ \slash p \, \gamma^\alpha \, 
  \frac{ \slash{k}_2 - \slash{\bar p}}{(k_2 - \bar p)^2} \, \gamma^\mu \, 
  \slash{\bar p} \, \gamma^\rho \, \frac{ \slash{k}_1 - \slash{\bar p}}{(k_1 - \bar p)^2} \,
  \gamma_\alpha \, \frac{\slash{p} + \slash{k}_1 - \slash{k}_2}{(p + k_1 - k_2)^2} \, 
  \gamma_\mu \, \frac{\slash{p} + \slash{k}_1}{(p + k_1)^2} \,
  \gamma_\rho \bigg] \, , \nonumber 
\eea
where $\omega = \sqrt{\hat{s}} \, (1 - z)$, and we have defined the integration measure
\beq 
  \int [dk_i] \, \equiv \, \frac{e^{\epsilon \gamma_{\rm E}}}{(4\pi)^{\epsilon}} \, 
  \mu_{\rm \overline{MS}}^{2 \epsilon} \, \int \frac{d^d k_i}{(2\pi)^d} \,\, ,
\label{defmeas}
\eeq 
with $d = 4 - 2 \epsilon$ and $\mu_{\rm \overline{MS}} = \mu \, e^{ - \gamma_{\rm E}/2} 
(4 \pi)^{1/2}$. One must then add to \eq{NNLO3}, and to all other contributions from the 
diagrams depicted in Fig.~\ref{sigbdiags}, similar terms obtained by interchanging
$p \leftrightarrow \bar{p}$ and by complex conjugation. Colour matrices have been 
neglected, given that we are focusing on the abelian-like part of the $K$-factor, which 
appears with an overall factor of $C_F^2$. 

In order to reproduce the NLP threshold logarithms in the $K$-factor, one must now 
classify all next-to-soft and collinear contributions. This is the subject of the following 
section.


\section{Method of regions analysis}
\label{sec:calc}

The method of regions is a systematic procedure for expanding loop integrals about 
their singular regions~\cite{Beneke:1997zp}, such that collinear and soft behaviours
are disentangled. Whilst a general proof of its validity is not yet available (see
for example~\cite{Pak:2010pt,Jantzen:2011nz}), it has been tested in a number
of highly non-trivial examples, most recently in Higgs production via gluon fusion 
at N$^3$LO~\cite{Anastasiou:2013mca}, a process closely related to Drell-Yan 
production. In what follows, however, we will apply the method of regions to identify 
all sources of NLP threshold logarithms, including next-to-soft contributions as well
as collinear ones, going beyond the purely soft or collinear limits considered in
Ref.~\cite{Anastasiou:2013mca}.

We begin by defining the directions collinear to the incoming quark and antiquark by 
the light-like vectors $n_+$ and $n_-$, defined such that $n_+^2 = n_-^2 = 0$ and 
$n_- \cdot n_+ = 2$.  The physical momenta (in the centre of mass frame) are related 
to these vectors via
\beq
  p^{\mu} \, = \, \frac{1}{2} \, (n_- p) \, n_+^\mu \, = \, 
  \frac{\sqrt{\hat s}}{2} \, n_+^{\mu} \, ,
  \qquad
  \bar p^{\mu} \, = \, \frac{1}{2} \, (n_+ \bar p) \, n_-^\mu \, = \, 
  \frac{\sqrt{\hat s}}{2} \, n_-^{\mu} \, ,
\label{scaling}
\eeq
where we introduced the short-hand notation $(n_{\pm} l) \equiv n_{\pm}^\mu l_\mu$. 
A generic momentum $l$ may then be decomposed into light-cone and transverse 
components according to
\beq
  l^{\mu} \, = \, \frac{1}{2} \, (n_- l) \, n_+^{\mu} + \frac{1}{2} \, (n_+ l) \,
  n_-^{\mu} + l_{\perp}^\mu \, .
\label{SCETdec}
\eeq
We now distinguish different regions for the momentum $l^\mu$ by the different scalings 
of its components, defined according to a book-keeping parameter $\lambda \sim 
\sqrt{E_{\rm soft}/E}$, where $E_{\rm soft} \sim \sqrt{\hat s} \, (1 - z)$, and  $E \sim 
\sqrt{\hat{s}}$ is the hard scale. More specifically, writing $l^\mu = \{ l_+, l_-, l_\perp \}$,
the relevant regions are defined as follows
\beqa
  & & {\rm Hard:} \quad l \sim \sqrt{\hat{s}} \left( 1, 1, 1 \right) \, ; \quad \;\;\;\;
  {\rm Soft:} \quad l \sim \sqrt{\hat{s}} \left( \lambda^2, \lambda^2, \lambda^2 \right) \, ;
  \nonumber  \\
  & & {\rm Collinear:} \quad l \sim \sqrt{\hat{s}} \left( 1, \lambda, \lambda^2 \right) \, ; \quad
  {\rm Anticollinear:} \quad l \sim \sqrt{\hat{s}} \left( \lambda^2, \lambda, 1 \right) \, .
\label{scalings}
\eeqa
In any given process, the external momenta are fixed. Here, for example, $p$ ($\bar{p}$) 
is by definition collinear (anticollinear), while $k_2$ is (next-to) soft. Different contributions 
to the $K$ factor then arise from various regions of the loop momentum $k_1$.

Our next task is to expand the propagators in \eq{NNLO3} in the different regions. 
Focusing, as an example, on those associated with the $p$ leg, the most complicated 
case is
\beq
  \frac{\slash{p} + \slash{k}_1 - \slash{k}_2}{(p + k_1 - k_2)^2} \, .
\label{comprop}
\eeq
Expanding to the second non-trivial order in $\lambda$ in the relevant momentum regions 
described above, this propagator becomes
\beqa
\label{col} 
  {\rm Hard}: \label{hard} & & \frac{\sqrt{\hat s} \, \frac{\slash{n}_+}{2} + 
  \slash{k}_1}{k_1^2 + (n_+ k_1) \,\sqrt{\hat s}} \, + \, \Bigg[ - \, \frac{\slash{k}_2}{k_1^2 + 
  (n_+ k_1) \sqrt{\hat s}} \nonumber \\
  & & \hspace{15mm} + \, \frac{ \left( (n_+ k_2) \, \sqrt{\hat s} + 2 (k_1 k_2) \right) 
  \left( \sqrt{\hat s} \, \frac{\slash{n}_+}{2} + \slash{k}_1 \right)}{(k_1^2 + (n_+ k_1) \,
  \sqrt{\hat s})^2} \Bigg] \, + \, \ord(\lambda^4) \, ;  \nonumber  \\
  {\rm Collinear:} &  & \frac{ \left( \sqrt{\hat s} + (n_- k_1) \right) \, \frac{\slash{n}_+}{2}}
  {k_1^2 + (n_+ k_1) \, \sqrt{\hat s} - (n_+ k_2) \left( \sqrt{\hat s} + (n_- k_1) \right)} 
  \nonumber \\ & & \hspace{15mm} + \Bigg[ \frac{\slash{k}_{1 \perp}}{k_1^2 + (n_+ k_1) \, 
  \sqrt{\hat s} - (n_+ k_2) \left( \sqrt{\hat s} + (n_- k_1) \right)} \nonumber \\ \nn
  & & \hspace{2cm} + \, \frac{ 2 \, (k_{1 \perp} k_{2 \perp}) \left( \sqrt{\hat s} + (n_- k_1) \right) 
  \frac{\slash{n}_+}{2}}{ \left( k_1^2 + (n_+ k_1) \, \sqrt{\hat s} - (n_+ k_2) 
  \left( \sqrt{\hat s} + (n_- k_1) \right) \right)^2} \Bigg] \nonumber \\
  & & \hspace{15mm} + \, \Bigg[ \frac{(n_+ k_1) \, \frac{n_-}{2} - \slash k_2}{k_1^2 + (n_+ k_1) \, 
  \sqrt{\hat s} - (n_+ k_2) \left( \sqrt{\hat s} + (n_- k_1) \right)} \\
  & & \hspace{2cm} + \, \frac{2 \, (k_{1\perp} k_{2\perp} ) \slash k_{1 \perp} + 
  \left( (n_+ k_1) \, (n_- k_2) - k_2^2 \right) \left( \sqrt{\hat s} + (n_- k_1) \right) \, 
  \frac{\slash{n}_+}{2}}{ \left( k_1^2 + (n_+ k_1) \, \sqrt{\hat s} - (n_+ k_2) 
  \left( \sqrt{\hat s} + (n_- k_1) \right) \right)^2} \nonumber \\ 
  & & \hspace{2cm} + \, \frac{4 \, (k_{1\perp} k_{2\perp})^2 \left( \sqrt{\hat s} + (n_- k_1) 
  \right) \frac{\slash{n}_+}{2}}{\left( k_1^2 + (n_+ k_1) \, \sqrt{\hat s} - (n_+ k_2) 
  \left( \sqrt{\hat s} + (n_- k_1) \right) \right)^3} \Bigg] + \ord( \lambda ) \, ; 
  \nonumber \\
  {\rm Anticollinear:} & & \frac{1}{\sqrt{\hat s}} \frac{\slash{n}_-}{2}
  + \frac{1}{(n_+ k_1)} \frac{\slash{n}_+}{2} + \Bigg[ 
  \frac{1}{(n_+ k_1)} \frac{\slash{k}_{1 \perp}}{\sqrt{\hat s}} \nonumber \\ 
  & & \hspace{5mm} + \, \left( - \, \frac{k_{1 \perp}^2}{(n_+ k_1)^2 \, \sqrt{\hat s}} + 
  \frac{(n_+ k_2)}{(n_+ k_1)^2}+ \frac{(n_- k_2)}{(n_+ k_1) \, \sqrt{\hat s}} \right)
  \frac{\slash{n}_+}{2} \Bigg] \nonumber \\ 
  & & \hspace{5mm} + \, \Bigg[ \left(- \, \frac{k_1^2}{(n_+ k_1) \, \hat s}
  + \frac{(n_+ k_2)}{(n_+ k_1) \, \sqrt{\hat s}} + \frac{(n_- k_2)}{\hat s}
  \right) \frac{\slash{n}_-}{2} - \frac{\slash{k}_2}{(n_+ k_1) \, \sqrt{\hat s}} \Bigg]
  + \ord(\lambda^3) \, ; \nonumber \\ 
  {\rm Soft:} & & \frac{1}{(n_+ k_1) - (n_+ k_2)} \, \frac{\slash{n}_+}{2}
  + \Bigg[ \frac{1}{(n_+ k_1) - (n_+ k_2)} \frac{\slash k_1 - \slash k_2}{\sqrt{\hat s}} 
  \nonumber \\ 
  & & \hspace{10mm} - \, \frac{1}{\big( (n_+ k_1) - (n_+ k_2) \big)^2} 
  \frac{(k_1 - k_2)^2}{\sqrt{\hat s}}  \frac{\slash{n}_+}{2} \Bigg] + \ord(\lambda^2) \, . 
  \nonumber
\eea
In order to clarify the power counting in \eq{col}, it may be useful
to note that the expansion of the propagator given in \eq{comprop} in
powers of $\lambda$ starts at ${\cal O}(\lambda^{0})$ in the hard and
in the anticollinear regions, while it starts at ${\cal
  O}(\lambda^{-2})$ in the collinear and soft regions. Moreover, the
Taylor expansion is in powers of $\lambda^2$ in the hard and soft
regions, while it is in powers of $\lambda$ in the collinear and
anticollinear regions. In \eq{col}, different orders in $\lambda$ are
enclosed within square brackets.

Similar expressions can be obtained for all other propagators, not all of which are 
independent (for example, the anticollinear region for the $p$ leg can be obtained 
from the collinear region on the $\bar{p}$ leg by relabelling $p \leftrightarrow \bar{p}$). 
After substituting all expanded propagators into \eq{NNLO3}, the integrals may be
carried out in dimensional regularisation using standard techniques. One may then 
repeat this procedure for the remaining diagrams in Fig~\ref{sigbdiags}. When this 
is done, it is useful to present results for two distinct sums of diagrams: those involving 
both quark legs, $p$ and $\bar{p}$, given in graphs (a)--(d) in Fig.~\ref{sigbdiags}, and 
those involving a single leg, given in graphs (e)--(h). Complete results to NLP accuracy 
are given below: for each region r, we write the $K$ factor as $K_{\rm r} (z) = K_{\rm E, \, 
r} (z) + K_{\rm NE, \, r} (z)$, separating the result into two parts, corresponding to leading
and next-to-leading order in the eikonal (soft) expansion of the amplitude in powers 
of $k_2$, before phase space integration. The NLP logarithms in the eikonal 
contributions $K^{(2)}_{\rm E, \, r} (z)$ arise exclusively from corrections to the 
eikonal phase space, as discussed in Ref.~\cite{Laenen:2010uz}. Next-to-eikonal
contributions $K^{(2)}_{\rm NE, \, r}$, on the other hand, consist of genuine corrections 
arising at the amplitude level.


\subsection{Hard region}
\label{hardreg}

After integration over the loop momentum $k_1$, and the real radiation phase space
for momentum $k_2$, we find that there is no contribution at LP or NLP arising from the 
hard region from diagrams (e)--(h). Diagrams (a)--(d), on the other hand, combine to give 
\beqa
\label{NNLOHard}
  K^{(2)}_{\rm E, \, h} (z) & = & \left( \frac{\as}{\pi} \right)^2 \,
  \Bigg[ \frac{2 {\cal D}_0(z)}{\eps^3} + \frac{- 4 + 3 {\cal D}_0 (z) - 
  4 {\cal D}_1 (z)}{\eps^2} \\
  & & \hspace{1.6cm} + \, \frac{- 6 + 8 {\cal D}_0 (z) - 6 {\cal D}_1 (z) + 4 {\cal D}_2 (z) 
  + 8 \log(1 - z)}{\eps} \nonumber \\
  & & \hspace{1.6cm}  - \, 16 + 16 {\cal D}_0 (z) - 16 {\cal D}_1 (z) + 6 {\cal D}_2 (z) - 
  \frac{8 {\cal D}_3 (z)}{3} \nonumber \\
  & & \hspace{1.6cm} + \, 12 \log(1 - z) - 8 \log^2(1 - z) \Bigg] \, , \nonumber \\
  K^{(2)}_{\rm NE, \, h} (z) & = & \left( \frac{\as}{\pi} \right)^2 \,
  \bigg[ - \frac{2}{\eps^3} + \frac{1 + 4 \log(1 - z)}{\eps^2} \\
  & & \hspace{1.6cm} + \, \frac{- 5 + 2 \log(1 - z) - 4 \log^2(1 - z)}{\eps} - 8 \nonumber \\ 
  & & \hspace{1.6cm} + \, 10 \log(1 - z) - 2 \log^2(1 - z) + \frac{8}{3} \log^3(1 - z) 
  \Bigg] \, . \nonumber 
\eeqa
In writing our results for $K$ factors, we have chosen $\mu^2_{\rm
  \overline{MS}} = q^2$, we have omitted the overall factor of
$C_F^2$, which is common to all our results, and we have also omitted,
for brevity, terms involving logarithms multiplied by transcendental
constants: these terms can easily be generated and do not carry any
new information.  Interestingly, we find that the plus distribution
terms in \eq{NNLOHard} suffice to reproduce all corresponding terms in
the exact NNLO Drell-Yan $K$-factor~\cite{Matsuura:1988sm}.  This
means that the remaining regions may not contribute any further LP
logarithms. We will briefly comment below on the interesting interplay
between soft and hard regions which is suggested by this result.


\subsection{Collinear and anticollinear regions}
\label{collreg}

By symmetry, the collinear and anticollinear regions must give the same contribution, 
after summing over all graphs in Fig.~\ref{sigbdiags}, and including those obtained via
complex conjugation and via the interchange $p \leftrightarrow \bar{p}$. The contribution 
from both regions from diagrams (a)--(d) is then
\beq
  K^{(2), \, {\rm a-d}}_{\rm NE, \, c + \bar c}(z) \, = \, \left( \frac{\as}{\pi} \right)^2 \, 
  \Bigg[ - \frac{1}{2 \eps^2} + \frac{3 \log(1 - z)}{2 \eps} + 1 - \frac{9}{4} \log^2(1-z) \Bigg] \, .  
\label{NNLOCollinearNEvert} 
\eeq
As expected, we find only a contribution starting at NE level. Note however that it is not 
true that individual diagrams have only next-to-soft contributions in the collinear region.
For example, diagrams (a), (c), (f) and (h) separately contain plus distribution terms. This,
however, is an artifact of having used the Feynman gauge, and eikonal terms cancel 
when diagrams are summed. Likewise, the contribution from diagrams (e)--(h) read
\beq
  K^{(2), \, {\rm e-h}}_{\rm NE, \, c + \bar c}(z) \, = \, \left( \frac{\as}{\pi} \right)^2 \, 
  \Bigg[ - \frac{1}{2 \eps^2} + \frac{- 5 + 6 \log(1 - z)}{4 \eps} - \frac{5}{2}  + 
  \frac{15}{4} \log(1-z) - \frac{9}{4} \log^2(1 - z) \Bigg] \, .  
\label{NNLOCollinearNEext} 
\eeq


\subsection{Soft region}
\label{softreg}

In this region, all integrals are scaleless, and thus vanish in
dimensional regularisation.  This is consistent with the fact that
eikonal terms have already been included in the hard region, according
to its definition in \eq{scalings}. So far as divergent terms are
concerned, this collocation of singular terms is not surprising: it is
well known that one can shift singularities from the IR to the UV by
using dimensional regularization as we have just done, taking
literally the vanishing of scaleless integrals without attempting to
distinguish the ultraviolet and the infrared singularities they
contain. It is interesting that, at least within the framework of a
method-of-regions analysis, this mechanism appears to extend to
finite, and even integrable, contributions to the cross section. Note
finally that this result is compatible with the approach taken in
Ref.~\cite{DelDuca:1990gz}, where the `hard' function is taken to
implicitly include the soft function, in order to extract the more
interesting collinear contributions.


\subsection{The complete abelian-like real-virtual NNLO $K$ factor}
\label{Ksum}

Combining results from the preceding subsections, the complete $K$ factor arising
from NNLO abelian-like real-virtual contributions, as computed by the method of regions,
is given by
\beqa
\label{K2tot}
  K^{(2)}_{\rm E + NE} (z) & = & \left( \frac{\alpha_s}{\pi} \right)^2 \Bigg[
  \frac{2 {\cal D}_0 (z) - 2} {\epsilon^3} + \frac{- 4{\cal D}_1 (z) + 3 {\cal D}_0 (z) +
  4 \log(1 - z) - 6}{\epsilon^2} \\ 
  & & \hspace{8mm} + \, \frac{ 16 {\cal D}_2 (z) -
  24 {\cal D}_1 (z) + 32 {\cal D}_0 (z)  - 16 \log^2 (1 - z) + 52 \log(1-z) - 49}{4 \epsilon} 
  \nonumber \\
  & & \hspace{8mm} - \, \frac{8 {\cal D}_3 (z)}{3} + 6 {\cal D}_2 (z) - 16 {\cal D}_1 (z) + 
  16 {\cal D}_0 (z) + \frac{8}{3} \log^3 (1 - z) \nonumber \\ 
  & & \hspace{8mm} - \frac{29}{2} \log^2 (1 - z) + \frac{103}{4} \log(1 - z) 
  - \frac{51}{2} \Bigg] \, . \nonumber 
\eeqa
We find that \eq{K2tot} reproduces exactly the result obtained in Ref.~\cite{Hamberg:1990np},
when the relevant diagrams are isolated\footnote{Note that separate results for the double-real   
  emission and for the real-virtual contribution to the NNLO Drell-Yan cross sections are not
  available in the literature: we have carried out an independent calculation of the relevant 
  diagrams~\cite{NEinprep}. One may furthermore verify that combining \eq{K2tot} with the 
  results of Ref.~\cite{Laenen:2010uz}, and with the appropriate mass-factorisation
  counterterms, reproduces the complete result of Ref.~\cite{Hamberg:1990np}.}, including
$z$-independent terms. This is an interesting fact: it reinforces the conjecture that one 
can carry out the calculation, either with the method of regions or in a factorized approach,
as a systematic expansion in powers of the distance from threshold, $1 - z$, including not
only functions that are (integrably) singular at threshold, but also polynomial dependence.

Whilst fully integrated results are useful in obtaining the final NLP threshold logarithms
in the $K$-factor, it is also useful to characterise what happens before the real emission
integration is carried out. This is the subject of the following section.


\section{Loop effects and the soft expansion}
\label{sec:loop}

In this section, we examine our results in light of the recently proposed next-to-soft theorems 
of Ref.~\cite{Cachazo:2014fwa,Casali:2014xpa}. In particular, we focus on the issue, pointed 
out in Ref.~\cite{Cachazo:2014dia}, and further discussed in Refs.~\cite{Bianchi:2014gla,
Bern:2014oka,He:2014bga}, that potential loop corrections to tree-level next-to-soft factors 
depend on the order in which the dimensional regularisation and soft expansions are carried 
out.

In \secn{sec:calc} we presented results for the hard, collinear and anticollinear regions, 
after the integration over the phase space of the real gluon (with momentum $k_2$) had 
already been performed. Implicit in the above calculation, but not immediately visible in 
the final result, is the fact that the different regions are weighted by different  scale-related 
factors. For example, after integration over $k_1$ (but before integration over $k_2$), 
the contribution from the {\it hard} region can be written schematically as
\beq
  {\rm Hard:} \quad  \frac{\left( 2 p \cdot \bar{p} \right)^{- \epsilon}}{\epsilon^2} \, 
  \Big[ {\rm E} + {\rm NE} + \ldots \Big] + \ord \left( \eps^{-1} \right) \, ,
\label{hardk1}
\eeq
where with E and NE we denote terms at $\ord (k_2^{-1})$ and $\ord (k_2^0)$ respectively,
and the ellipsis denotes higher-order terms in the soft expansion. Likewise, the collinear 
region contributes terms of the form\footnote{As mentioned above, the presence of NE 
  terms only in \eq{colk1} is the effect of a cancellation of eikonal terms which are present 
  in individual diagrams.}
\beq
  {\rm Collinear:} \quad \frac{ \left(- 2 p \cdot k_2 \right)^{- \epsilon}}{\epsilon} \, 
  \Big[ {\rm NE} + \ldots \Big] + \ord \left( \eps^0 \right) \, ,
\label{colk1}
\eeq
while the anticollinear region is naturally obtained by replacing $p$ with $\bar{p}$. That 
these particular scales arise is not surprising: they are the only scales that survive in 
each given region. It is now clear why the eikonal terms are reproduced from the hard 
region in this formalism: these terms must arise from the soft-collinear factorisation 
formula, in which the relevant hard, soft and jet functions cannot depend on the scales
$(p \cdot k_2)$ and $(\bar{p}\cdot k_2)$, as they are defined without reference to an 
extra emission. Interestingly, the collinear regions depends on $z$ through
\beq
  \left( - 2 p \cdot k_2 \right)^{- \epsilon} \, \sim \, (1 - z)^{ - \epsilon} \, ,
\label{collz}
\eeq
and the same dependence arises in the anti collinear region, through
\beq
  \left( - 2 \bar{p} \cdot k_2 \right)^{- \epsilon} \, \sim \, (1 - z)^{- \epsilon} \, .
\label{acollz}
\eeq
This dependence is responsible for the pattern of NLP threshold logarithms in 
Eqs.~(\ref{NNLOCollinearNEvert}) and (\ref{NNLOCollinearNEext}), which is generated 
as follows. The phase space for the real gluon emission contains a further $z$ dependent
factor $[(1 - z)/z]^{1 - 2 \epsilon}$ (see for example~\cite{Hamberg:1990np}), so that the 
$k_2$ integration leads to a result of the form
\beq
  \frac{(1 - z)^{- 3 \epsilon}}{\epsilon^2} \, = \, \frac{1}{\epsilon^2} - 
  \frac{3}{2} \frac{\log (1 - z)}{\epsilon} + \frac{9}{2} \log^2 (1 - z) \, ,
\label{logexp}
\eeq
where the additional power of $\epsilon^{-1}$ arises after carrying out the phase space 
integration. \eq{logexp} carries exactly the pattern of NLP threshold logarithms observed 
in Eqs.~(\ref{NNLOCollinearNEvert})  and (\ref{NNLOCollinearNEext}), after multiplying 
by the appropriate normalisation. It is clear that terms proportional to $(p \cdot k_2)^{-
\epsilon}$ play a crucial role in order to correctly reproduce the known Drell-Yan $K$-factor
at NLP accuracy.

The factor $(p\cdot k_2)^{-\epsilon}$ is very interesting from the point of view of the 
soft expansion in powers of $k_2$. Such a factor would be absent if one performed the 
soft expansion {\it before} the dimensional regularisation expansion, and it is clear from 
individual Feynman diagrams such as that of \eq{NNLO3} why this is the case: carrying 
out the soft expansion before the $\epsilon$ expansion amounts to expanding the 
integrand before integration over the virtual momentum $k_1$. This, for example, 
replaces the mixed denominator according to
\beq
  \frac{1}{(p - k_1 - k_2)^2} \, \rightarrow \, \frac{1}{(p - k_1)^2} \, ,
\label{denreplace} 
\eeq so that logarithmic dependence on $(p \cdot k_2)$ can no longer
occur in the final result: only logarithmic dependence on $p \cdot
\bar{p}$, which is still present as a scale in the denominator, can
arise. This observation fixes the order in which these expansions must
be carried out: to get the right answer, one must integrate over
virtual momenta before expanding in soft momentum\footnote{More
  precisely, one is allowed to neglect terms proportional to $k_2$ in
  the numerators of loop integrands. One must, however, keep
  denominators intact, since they can lead to logarithmic
  dependence.}. Note that the only terms which are ``problematic''
from the point of the view of the soft expansion ({\it i.e.}  that
depend on the sequential order of the soft and $\epsilon$ expansions)
are those involving overall powers of $(p \cdot k_2)^{- \epsilon}$ or
$(\bar{p} \cdot k_2)^{- \epsilon}$.  These arise exclusively from the
(anti-)collinear regions, which is not surprising: in the hard region,
one may neglect the scales $(p \cdot k_2)$ and $(\bar{p} \cdot k_2)$
with respect to the hard scale $p \cdot \bar{p}$, leading to a
power-like suppression of next-to-soft effects. That the collinear
region leads to a breakdown of the Low-Burnett-Kroll
theorem~\cite{Low:1958sn,Burnett:1967km}, due to the absence of a hard
scale, is well-known, and was first pointed out by Del
Duca~\cite{DelDuca:1990gz}. It can also be understood from an
effective field theory point of view~\cite{Duffinprep}. Furthermore,
the need to first perform the dimensional regularisation expansion has
been recently discussed in the Ref.~\cite{Bern:2014oka}. Here, though,
we see a concrete example of the impact of this effect on any
systematic treatment of threshold corrections.


\section{Discussion}
\label{sec:conclude}

In this paper, we have performed a case study of threshold effects in
Drell-Yan production at next-to-leading power. We focused in
particular on reproducing known logarithmic contributions to the
real-virtual part of the NNLO $K$-factor, from the point of view of a
threshold expansion: this is the first order at which there is an
interplay between real and virtual gluons, so that collinear
singularities may interfere with the soft expansion. As a consequence,
our study allowed us also to investigate potential loop corrections to
recently proposed next-to-soft
theorems~\cite{Cachazo:2014fwa,Casali:2014xpa}. Our main goal,
however, is to provide useful data for the development of a generally
applicable resummation formalism for NLP threshold logarithms,
building on previous efforts~\cite{Grunberg:2009yi,
  Dokshitzer:2005bf,Laenen:2008ux,Almasy:2010wn,Laenen:2008gt,Laenen:2010uz,Ball:2013bra}.
We used the method of
regions~\cite{Beneke:1997zp,Pak:2010pt,Jantzen:2011nz} to separate out
contributions from the hard, soft and (anti)-collinear momentum
configurations. A first gratifying result is that a systematic
application of this method beyond leading power allowed us to
reproduce exactly all corresponding terms in the exact calculation,
including $z$-independent contributions. This confirms that all
threshold logarithms to this accuracy arise from soft or collinear
singularities, and reinforces the idea of using the threshold
expansion as a systematic tool for the analysis of QCD cross sections,
both at finite orders~\cite{Anastasiou:2013mca, deFlorian:2014vta} and
in the context of threshold resummation. Our analysis also shows that
collinear regions contribute logarithmic dependence on soft momenta,
which affects the NLP threshold logarithms one obtains after
integration over the real gluon phase space. This fixes the order in
which the dimensional regularisation and soft expansions must be
carried out, as was also discussed in
Refs.~\cite{DelDuca:1990gz,Cachazo:2014dia,Bianchi:2014gla,Bern:2014oka,He:2014bga}.
It is instructive and useful to see exactly how this mechanism
operates in the familiar context of Drell-Yan production.

Our results will be instrumental in the construction of a systematic all-order treatment
of threshold effects at NLP accuracy: they carry the basic information that the interplay 
between soft and collinear effects is considerably more intricate at NLP than it is in
standard leading-power soft-collinear factorization. A systematic treatment will require
the introduction of new operator matrix elements, incorporating the effects of 
non-factorizing soft radiation from collinearly enhanced configurations, as first
suggested in Ref.~\cite{DelDuca:1990gz}. Work to implement these considerations
in a systematic way, beginning with the relatively simple case of electroweak annihilation 
processes, is in progress.


\section*{Acknowledgments}

We thank Babis Anastasiou, Pietro Falgari, Einan Gardi, Duff Neill and Marco 
Volponi for useful discussions and correspondence. This work was supported 
by the Research Executive Agency (REA) of the European Union through the 
contracts PITN-GA-2010-264564 (LHCPhenoNet) and PITN-GA-2012-316704 
(HIGGSTOOLS), by MIUR (Italy), under contract 2010YJ2NYW$\_$006, and 
by the University of Torino and the Compagnia di San Paolo under contract 
ORTO11TPXK. DB and EL have been supported by the Netherlands
Foundation for Fundamental Research of Matter (FOM) programme 104,
``Theoretical Particle Physics in the Era of the LHC'', and by the National 
Organization for Scientific Research (NWO). CDW is supported by the UK 
Science and Technology Facilities Council (STFC). We are grateful to the 
Higgs Centre for Theoretical Physics at the University of Edinburgh, where 
part of this work was carried out, for warm hospitality. EL thanks the Galileo 
Galilei Institute for Theoretical Physics for hospitality, and the INFN for partial 
support during the completion of this work.




\bibliography{refs.bib}


\end{document}